\documentclass{article}
\usepackage{cite}
\usepackage{amsmath,amssymb,amsfonts}
\usepackage{algorithmic}
\usepackage{graphicx}
\usepackage{textcomp}
\usepackage{ulem}

\begin{document}
\title{Conditions for domain-free negative capacitance}
\author{Prasanna Venkatesan Ravindran, Priyankka Gundlapudi \\Ravikumar, Asif Islam Khan}

\date{}
\maketitle

\begin{abstract}
While negative capacitance has been demonstrated in ferroelectric-dielectric heterostructures in the form of capacitance enhancement, all experimental evidence, to date, suggests the existence of domains therein.  Here, we address the question: what are the conditions to achieve ideal, domain-free negative capacitance  in ferroelectric-dielectric heterostructures? Our main claim is  that for given thicknesses of the ferroelectric and the dielectric layers, there is a critical value of domain wall energy parameter ---  above which the system would be stabilized in an ideal and robust domain-free negative capacitance state and would be robust against domain formation.  Our analyses suggest that to achieve ideal negative capacitance, efforts should lie in understanding the means to control the domain wall energy on all fronts, both theory and experiments via high throughput design, discovery, and engineering of ferroelectrics. 
\end{abstract}

\section{Introduction}
\label{sec:introduction}
Negative capacitance (NC) is an unstable state in a ferroelectric material in which the changes in the dielectric polarization ($P$) and the electric field ($E$) can occur in opposite directions.\cite{salahuddin2008use} Based on the Landau-Devonshire theory of ferroelectricity, a ferroelectric negative capacitance state  can be stabilized in a ferroelectric-dielectric heterostructure. When a ferroelectric thickness $t_{F}$ is smaller than a critical value $t_{F,c}$\ for a given dielectric thickness $t_{D}$, the depolarizing field  due to the dielectric layer can drive the ferroelectric layer into a homogeneous, zero polarization ($P=0$), negative capacitance state {(Fig. 1)}. Such a state minimizes the free energy of the FE-DE heterostructure. In doing so, the ferroelectric layer can passively amplify the voltage at the FE-DE interface and enhance the capacitance of the composite structure above that of the constituent dielectric layer. However, it has long been known  that such a heterostructure can also exist in a multi-domain, ferroelectric state.\cite{lines2001principles, bratkovsky2006depolarizing, kopal1997domain, kopal1999displacements, zubko2010x} One of the early debates on this topic has been whether the formation of a domain pattern precludes the negative capacitance response from the ferroelectric layer. This debate was put to rest after Zubko et al.\cite{zubko2016negative} experimentally demonstrated capacitance enhancement in a multi-domain, perovskite-based FE-DE heterostructure, followed by observation of local regions of negative dielectric permittivity in polar vortices and skyrmions\cite{yadav2019spatially,das2021local}.

\begin{figure}[!b]
 \center
   \includegraphics[width=0.7\textwidth]{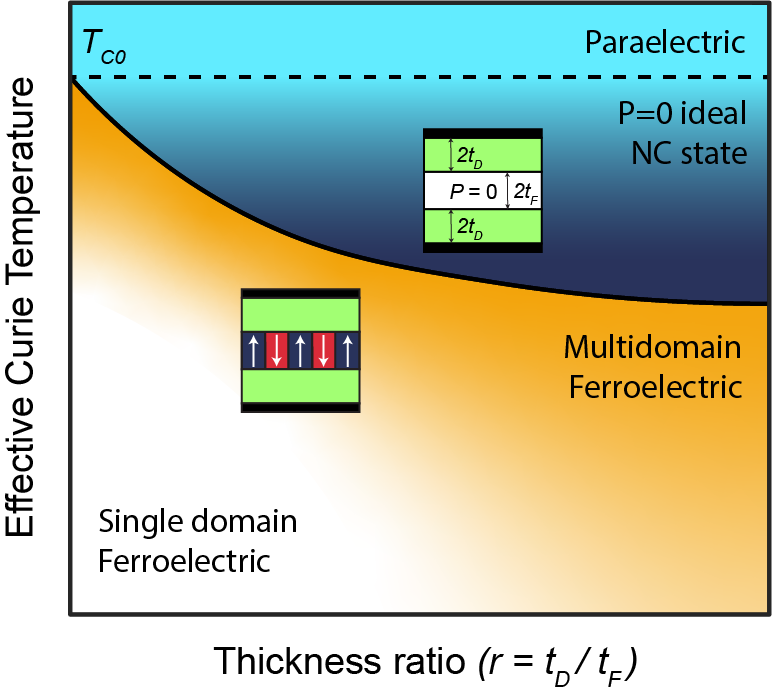}
   \caption{Phase plot of effective Curie temperature as a function of the thickness ratio $r$: The black curve shows the transition from a homogeneous zero-polarization, negative capacitance state to a multi-domain ferroelectric state in a FE-DE heterostructure. Despite several experimental efforts, the former has remained elusive leading to questions about the existence of such an ideal negative capacitance state. This letter analyzes the conditions on the domain wall energy parameter $D$ for which the zero-polarization state becomes energetically more favorable than the multi-domain state at a given operating temperature. $T_{C0}$ is the Curie temperature of the standalone ferroelectric material above which the material is paraelectric. $t_{D}$ and $t_F$ stand for the dielectric and ferroelectric thickness, respectively, while $P$ represents the polarization. }\label{intro}
 \end{figure}

That being said, the answer to the question, whether a ferroelectric layer can exist in a globally stabilized, zero polarization ($P=0$), negative capacitance state in a FE-DE structure lacks clarity. Such a state has neither been experimentally demonstrated nor been studied theoretically in sufficient detail. Negative capacitance in a multi-domain scenario has recently been theoretically analyzed in Ref. \cite{park2019modeling, saha2020multi, hoffmann2018stabilization, cano2010multidomain}. Hoffmann \textit{et al.} showed a homogeneous, $P=0$, NC state can exist if the lateral dimensions of the heterostructure is constrained below a critical limit.\cite{hoffmann2018stabilization} Rollo \textit{et al.} presented a case for the stabilization of negative capacitance in MFDM structures, emphasizing the importance of the quality of the ferroelectric-dielectric interface.\cite{rollo2020stabilization} Despite these works making earnest attempts to address the problem, the integration of the Landau-Ginzburg-Devonshire framework with the electrostatics of ferroelectric-dielectric heterostructures is known for mathematical inconsistencies stemming from models not being self-consistent. Phase-field simulations have been used to address these inconsistencies and Saha \textit{et al.} took into account the spatial variation of polarization within a domain and explained the crucial role of domain wall energy in stabilizing a homogeneous, zero polarization state ($P=0$) with negative capacitance in such heterostructures using phase-field simulations. \cite{saha2020multi}

In this work, we address this question by using a model for a multi-domain FE-DE heterostructure presented in \cite{stephanovich2005domain} based on the electrostatics of the multi-domain state at temperatures close to the Curie temperature $T_{C0}$, where the heterostructure transitions from a ferroelectric to a paraelectric phase. In FE-DE heterostructures, the effective Curie temperature $T_C'$ decreases as the volume fraction of the dielectric layer is increased. For a given heterostructure, at temperatures between $T_{C0}$ and $T_C'$, the homogeneous, zero polarization, negative capacitance state is stabilized before the heterostructure transitions to a multi-domain ferroelectric state below $T_C'$. (Fig. \ref{intro}) Our key result is that for a given FE-DE heterostructure with ferroelectric thickness $t_{F}<t_{F,C}$ and operating temperature $T_{op}$, there exists a critical domain wall energy required to stabilize the ferroelectric layer in this zero polarization, negative capacitance state. Above this critical energy, the system is bound to be stable in this state irrespective of the length scale of the sample in the lateral dimensions. 
\vspace{-0.1in}

\section{Single domain scenario in ferroelectric-dielectric heterostructure}
 
We first review the concept of the critical ferroelectric thickness $t_{F,C}$ in a FE-DE heterostructure in the single domain scenario as originally described in Ref. \cite{salahuddin2008use}. The free energy density per unit volume of a ferroelectric $U_{F}$ can be expressed as an even order polynomial of $P$--$i.e.$, $U_{F}=\sum \limits_{i=1,2...} \alpha_i P^{2\cdot i}$ where $\alpha_i$ are anisotropy constants. For the sake of simplicity, we assume that the ferroelectric material exhibits a first order phase transition--$i.e.$, $\alpha_1 = 1/2\epsilon_F(T)$ where $\epsilon_F = M/(T - T_{C0})$ based on the Curie-Weiss law and $\alpha_i>0$ for $i>1$. Here, $M$ and $T_{C0}$ are the Curie constant and the Curie temperature, respectively. The free energy per unit volume of dielectric layer is represented by a single well energy landscape $U_{D}=D_D^2/2\epsilon_{D}$, with  $D_D$ and $\epsilon_{D}$  being the electric flux and the dielectric permittivity of the dielectric layer. Equating $D_{D}$ to $P$, the energy density of the heterostructure is given by:
\begin{eqnarray}
    U&=&\frac{t_FU_F+t_DU_D}{t_F+t_D}=\frac{1}{2\epsilon'(T)}P^2+ \sum \limits_{i=2,...} \alpha_i P^{2\cdot i}\label{U}
\end{eqnarray}
\noindent where $\epsilon'$ is the effective dielectric constant of the heterostructure. $\epsilon'$ has the same functional form as that of $\epsilon_F$, with $T_C$ replaced by an effective Curie temperature $T_C'$, given as 

\begin{equation}
    T_C'=T_C-\frac{M}{\epsilon_D}r  \label{Tc_}
\end{equation} 

\noindent where, $r$ is the ratio between the thicknesses of the dielectric and the ferroelectric layers ($r={t_D}/{t_F}$). At an operating temperature $T_{op}$, the $P = 0$ state will be stable if $T_C' < T_{op} < T_{C0}$. For a given $\alpha_i$ and $T_{C0} > T_{op}$, the effective Curie temperature is lower than the operating temperature if $t_F < t_{F,C} = Mt_D/(T_{C0} - T_{op})\epsilon_D$. This implies that for $t_F < t_{F,C}$, the ferroelectric will be stable in the $P = 0$ state.

\section{Multi-domain scenario in ferroelectric-dielectric heterostructure}

We now consider the multi-domain scenario. We assume that each FE layer of thickness $t_{F}$ is sandwiched between two DE layers each with a thickness of $t_{D}$. The electrostatics of this system close to the transition temperature has been adapted from \cite{stephanovich2005domain}. {We modify the Eq. 2 in \cite{stephanovich2005domain} to introduce a unit-less proportionality constant, the domain wall energy parameter $D$ as a measure of the domain wall energy in the multi-domain state resulting in the domain wall energy term being $D^2\xi_0^2 \nabla^2 P$. A higher $D$ implies a higher domain wall energy and a higher threshold to forming domains while a lower one corresponds to a lower domain wall energy which would make domain formation easier.} Taking into consideration the domain wall energy, the $z$ component of the polarization in the ferroelectric layer $P$ is given by the non-linear Ginzburg-Landau equation:
\begin{equation}
    (T/T_{C0} - 1) P + P_0^{-2}P^3 - D^2 \xi_0^2 \nabla^2 P = \frac{\epsilon_{\parallel}}{4\pi}E_z^{(f)} \label{MD}
\end{equation}

\begin{table}[!t]
\caption{Material parameters for the ferroelectric and the dielectric layers}
\begin{center}
\begin{tabular}{cc}\hline
Parameter   & Value\\\hline
\textbf{Dielectric layer }& \\
Dielectric permittivity $\epsilon_{D}$ & 300$\epsilon_\circ$\\
\textbf{Ferroelectric layer}& \\
Out-of-plane permittivity $\epsilon_{\perp}$ & 120$\epsilon_\circ$ \\
Curie temperature $T_{C0}$ & 1244 K \\
Curie constant $M$ & 4.1$\times$10$^5$ K $\times \epsilon_\circ$\\
In-plane permittivity $\epsilon_{\parallel}$ & $M$/$T_{C0}$ \\
Domain wall half width $\xi_0$ & 0.5 nm\\
\hline
\end{tabular}
\end{center}
\label{table 1}
\vspace{-0.2in}
\end{table}

\noindent Following a similar approach, we obtain the effective Curie temperature $T_C'$ of the FE-DE heterostructures numerically. The effective Curie temperature is defined as the highest temperature at which a domain width(in turn, a domain configuration) is defined.

The FE and DE material parameters as used in this work are based on a PbTiO$_3$/SrTiO$_3$ heterostructure and are listed in the table \ref{table 1}.\cite{zubko2016negative} Over the last decade, works have reported perovskite-structured FE-DE superlattices with individual ferroelectric layers as small as a few unit cells.\cite{zubko2016negative,das2021local,yadav2019spatially} Simultaneously, perovskite structure ferroelectrics also have been scaled down to below 10 nm.\cite{pertsev2003coercive,kim2005critical} While the model is valid for any thickness range, in this work, we consider FE-DE superlattices with alternating layers of ferroelectric and dielectric with $t_F= 5$ nm.

\begin{figure}[t]
\center
\includegraphics[width=0.75\textwidth]{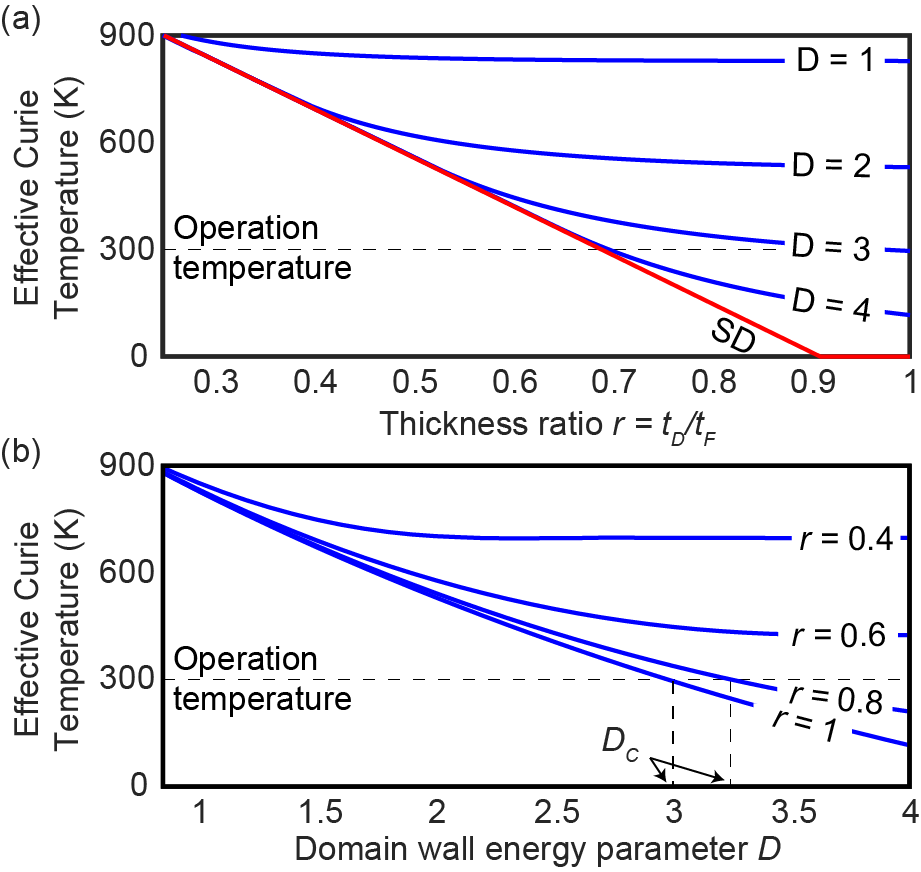}\\
\caption{The variation of effective Curie temperature (a) with thickness ratios for different $D$ and (b) with domain wall energy parameters for different $r$ in FE-DE heterostructures with $t_F$ = 5 nm. {As $D$ is increased, the effective Curie temperatures of multidomain heterostructures(blue) approach the effective Curie temperatures calculated from the single domain, homogeneous polarization model(red).}
}
\label{variation}
\end{figure}

For a fixed domain wall energy parameter $D$, the variation of the effective Curie temperature with the thickness ratio of dielectric to the ferroelectric is shown in Fig. \ref{variation}(a). For lower values of $D$, the $T_C'$ saturates closer to the $T_{C0}$. As $D$ is increased, the curves approach the behavior of a domain-free ferroelectric system{(red curve)} as domain formation is made unfavorable by the increasing domain wall energy. The operating temperature is assumed to be the 300 K in this work. The decreasing $T_C'$ with increasing $D$ shows that for suitable $D$, the FE-DE structure would be stabilized in a $P=0$, negative capacitance state at the operating temperature. Fig. \ref{variation}(b) shows variation of $T_C'$ with the $D$ for different dielectric thicknesses. We observe that $T_C'$ decreases with increasing $D$. For $t_F < t_{F,C}$(equivalently, $r > r_C$), there is critical domain wall energy parameter $D_C$ at which $T_C' < T_{op}$  indicating that the ferroelectric will be stabilized in a zero-polarization, negative capacitance state as suggested in Ref. \cite{salahuddin2008use} at room temperature for $D \ge D_C$.

\begin{figure}[]
\center
\includegraphics[width=0.75\textwidth]{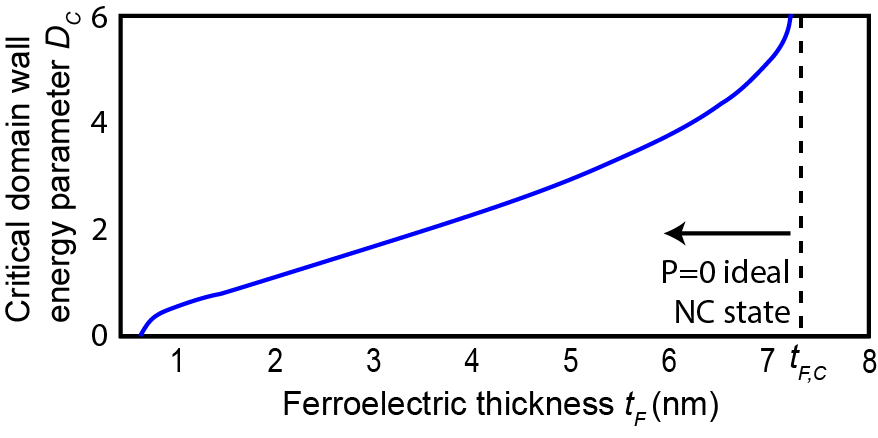}\\
\caption{For a given $t_{F}<t_{F,C}$ there exists a critical domain wall energy parameter $D_C$ above which the system is in a stable zero-polarization negative capacitance state at an operating temperature of 300 K and for $t_D$ = 5 nm. While $D$ is an intrinsic property, the critical domain wall energy parameter $D_C$ depends on the thickness of the ferroelectric and is not an intrinsic property.
}
\label{critical}
\end{figure}

The critical domain wall parameter $D_C$ is a function of the dielectric properties of the ferroelectric material and the ferroelectric and dielectric thicknesses. Based on the material parameters, we obtain $t_{F,c}=Mt_D/(T_{C0} - T_{op})\epsilon_D = 7.24$ nm. For a given $t_D$ and $t_F < t_{F,C}$, there exists a minimum domain wall energy required to stabilize the ferroelectric in a single domain, homogeneous $P = 0$, negative capacitance state. Fig. \ref{critical} shows the variation of the critical domain wall energy parameter $D_C$ with the ferroelectric thickness $t_F$ for $t_F<t_{F,C}$.

{Recent studies have reported that HfO$2$ exhibits a negative domain wall energy ($f_{dw}= -18$ mJ/m$^2$) for 180$^\circ$ domains, making the formation of these domains infinitely easy.\cite{lee2020scale} However, since $D_C$ is always positive, it is not possible to stabilize HfO$_2$ layers dominated by these domain walls in a $P=0$ ideal negative capacitance state in a FE-DE system, regardless of the dielectric and ferroelectric thickness. Nevertheless, negative capacitance effects have been experimentally observed in fluorite-type ferroelectric systems such as HfO$_2$ in previous studies, indicating that the current understanding of these materials may be insufficient for exploring negative capacitance at the nanoscale. \cite{zhou2016ferroelectric,li2018negative,cheema2022ultrathin} Moreover, practical HfO$_2$ contains multiple domains with about ten types of 180$^\circ$ and 90$^\circ$ domain walls within a single grain, which presents exciting opportunities for exploring the physics of negative capacitance in these materials.\cite{grimley2018atomic,hoffmann2021progress,lee2021domains}}

{Research focused on  domain wall energies and domain wall widths for different materials in different superlattices and under different stress and mechanical boundary conditions can offer valuable insights into the pathways for stabilizing the single-domain zero-polarization negative capacitance state. For example, HfO$_2$ has a domain wall energy of -18 mJ/m$^2$ and a domain wall width of 2.5 \text{\AA} \cite{lee2020scale} while strained PbTiO$_3$ on SrTiO$_3$ substrates and KNO$_3$ on KTaO$_3$ substrates have a domain wall width of 1 nm and 1.2 nm respectively at low-temperatures.\cite{zubko2016negative,stephanovich2005domain} However, this database is currently limited, and expanding it to encompass a broader range of materials would be a valuable pursuit.}

\section{Conclusion}

In conclusion, we have analyzed the possibility of a zero-polarization, negative capacitance state in ferroelectric-dielectric heterostructures using numerical results from a multi-domain model of the heterostructure. We show that for a given thickness of the dielectric and ferroelectric layers, there exists a critical value of the domain wall energy parameter only above which a single-domain zero-polarization, ideal negative capacitance state becomes stable. This points towards the need for further exploration of the energetics of domain walls and other functional materials like antiferroelectrics to achieve ideal stabilized negative capacitance that can potentially lead to ultra-low power transistors.

\section{Acknowledgements}
The work was supported by the National Science Foundation (NSF award no. 2047880 and 1810005). P.V.R. was supported by CRNCH PhD fellowship at Georgia Tech.

\bibliographystyle{IEEEtran}
\bibliography{references}

@article{stephanovich2005domain,
  title={Domain-enhanced interlayer coupling in ferroelectric/paraelectric superlattices},
  author={Stephanovich, VA and Luk’yanchuk, IA and Karkut, MG},
  journal={Physical review letters},
  volume={94},
  number={4},
  pages={047601},
  year={2005},
  publisher={APS}
}

@article{salahuddin2008use,
  title={Use of negative capacitance to provide voltage amplification for low power nanoscale devices},
  author={Salahuddin, Sayeef and Datta, Supriyo},
  journal={Nano letters},
  volume={8},
  number={2},
  pages={405--410},
  year={2008},
  publisher={ACS Publications}
}

@article{zubko2016negative,
  title={Negative capacitance in multidomain ferroelectric superlattices},
  author={Zubko, Pavlo and Wojde{\l}, Jacek C and Hadjimichael, Marios and Fernandez-Pena, St{\'e}phanie and Sen{\'e}, Ana{\"\i}s and Luk’yanchuk, Igor and Triscone, Jean-Marc and {\'I}{\~n}iguez, Jorge},
  journal={Nature},
  volume={534},
  number={7608},
  pages={524--528},
  year={2016},
  publisher={Nature Publishing Group}
}

@article{lee2020scale,
  title={Scale-free ferroelectricity induced by flat phonon bands in HfO2},
  author={Lee, Hyun-Jae and Lee, Minseong and Lee, Kyoungjun and Jo, Jinhyeong and Yang, Hyemi and Kim, Yungyeom and Chae, Seung Chul and Waghmare, Umesh and Lee, Jun Hee},
  journal={Science},
  volume={369},
  number={6509},
  pages={1343--1347},
  year={2020},
  publisher={American Association for the Advancement of Science}
}

@article{saha2020multi,
  title={Multi-Domain Negative Capacitance Effects in Metal-Ferroelectric-Insulator-Semiconductor/Metal Stacks: A Phase-field Simulation Based Study},
  author={Saha, Atanu K and Gupta, Sumeet K},
  journal={Scientific reports},
  volume={10},
  number={1},
  pages={1--12},
  year={2020},
  publisher={Nature Publishing Group}
}

@article{yadav2019spatially,
  title={Spatially resolved steady-state negative capacitance},
  author={Yadav, Ajay K and Nguyen, Kayla X and Hong, Zijian and Garc{\'\i}a-Fern{\'a}ndez, Pablo and Aguado-Puente, Pablo and Nelson, Christopher T and Das, Sujit and Prasad, Bhagwati and Kwon, Daewoong and Cheema, Suraj and Khan, Asif I. and Hu, Chenming and {\'I}{\~n}iguez, Jorge and Junquera, Javier and Chen, Long-Qing and Muller, David A. and Ramamoorthy, Ramesh and Salahuddin, Sayeef},
  journal={Nature},
  volume={565},
  number={7740},
  pages={468--471},
  year={2019},
  publisher={Nature Publishing Group}
}

@article{park2019modeling,
  title={Modeling of negative capacitance in ferroelectric thin films},
  author={Park, Hyeon Woo and Roh, Jangho and Lee, Yong Bin and Hwang, Cheol Seong},
  journal={Advanced Materials},
  volume={31},
  number={32},
  pages={1805266},
  year={2019},
  publisher={Wiley Online Library}
}

@article{hoffmann2018stabilization,
  title={On the stabilization of ferroelectric negative capacitance in nanoscale devices},
  author={Hoffmann, Michael and Pe{\v{s}}i{\'c}, Milan and Slesazeck, Stefan and Schroeder, Uwe and Mikolajick, Thomas},
  journal={Nanoscale},
  volume={10},
  number={23},
  pages={10891--10899},
  year={2018},
  publisher={Royal Society of Chemistry}
}

@book{lines2001principles,
  title={Principles and applications of ferroelectrics and related materials},
  author={Lines, Malcolm E and Glass, Alastair M},
  year={2001},
  publisher={Oxford university press}
}

@article{bratkovsky2006depolarizing,
  title={Depolarizing field and “real” hysteresis loops in nanometer-scale ferroelectric films},
  author={Bratkovsky, AM and Levanyuk, AP},
  journal={Applied physics letters},
  volume={89},
  number={25},
  pages={253108},
  year={2006},
  publisher={American Institute of Physics}
}

@article{kopal1997domain,
  title={Domain formation in thin ferroelectric films: The role of depolarization energy},
  author={Kopal, Anton{\'\i}n and Bahn{\'\i}k, Tom{\'a}{\v{s}} and Fousek, Jan},
  journal={Ferroelectrics},
  volume={202},
  number={1},
  pages={267--274},
  year={1997},
  publisher={Taylor \& Francis}
}

@article{kopal1999displacements,
  title={Displacements of 180 domain walls in electroded ferroelectric single crystals: The effect of surface layers on restoring force},
  author={Kopal, A and Mokr{\`y}, P and Fousek, J and Bahnik, T},
  journal={Ferroelectrics},
  volume={223},
  number={1},
  pages={127--134},
  year={1999},
  publisher={Taylor \& Francis}
}

@article{zubko2010x,
  title={{X-ray diffraction studies of 180$^{\circ}$ ferroelectric domains in PbTiO$_3$/SrTiO$_3$ superlattices under an applied electric field}},
  author={Zubko, Pavlo and Stucki, Nicolas and Lichtensteiger, C{\'e}line and Triscone, J-M},
  journal={Physical review letters},
  volume={104},
  number={18},
  pages={187601},
  year={2010},
  publisher={APS}
}

@article{cano2010multidomain,
  title={Multidomain ferroelectricity as a limiting factor for voltage amplification in ferroelectric field-effect transistors},
  author={Cano, A and Jim{\'e}nez, D},
  journal={Applied Physics Letters},
  volume={97},
  number={13},
  pages={133509},
  year={2010},
  publisher={American Institute of Physics}
}

@article{das2021local,
  title={Local negative permittivity and topological phase transition in polar skyrmions},
  author={Das, Sujit and Hong, Zijian and Stoica, VA and Gon{\c{c}}alves, MAP and Shao, Yu-Tsun and Parsonnet, Eric and Marksz, Eric J and Saremi, Sahar and McCarter, MR and Reynoso, A and Long, C.J. and Hagerstrom, AM and Meyers, D and Ravi, V and Prasad, B and Zhou, H and Zhang, Z and Wen, H and G{\'o}mez-Ortiz, F and Garc{\'i}a-Fern{\'a}ndez, P and Bokor, J and {\'I}{\~n}iguez, J and Freeland, J. W. and Orloff, N. D. and Junquera, J and Chen, L.-Q. and Salahuddin, S and Muller, D. A. and Martin, L.W. and Ramesh, R},
  journal={Nature materials},
  volume={20},
  number={2},
  pages={194--201},
  year={2021},
  publisher={Nature Publishing Group}
}

@article{grimley2018atomic,
  title={Atomic structure of domain and interphase boundaries in ferroelectric HfO2},
  author={Grimley, Everett D and Schenk, Tony and Mikolajick, Thomas and Schroeder, Uwe and LeBeau, James M},
  journal={Advanced Materials Interfaces},
  volume={5},
  number={5},
  pages={1701258},
  year={2018},
  publisher={Wiley Online Library}
}

@article{hoffmann2021progress,
  title={Progress and future prospects of negative capacitance electronics: A materials perspective},
  author={Hoffmann, Michael and Slesazeck, Stefan and Mikolajick, Thomas},
  journal={APL Materials},
  volume={9},
  number={2},
  pages={020902},
  year={2021},
  publisher={AIP Publishing LLC}
}

@article{lee2021domains,
  title={Domains and domain dynamics in fluorite-structured ferroelectrics},
  author={Lee, Dong Hyun and Lee, Younghwan and Yang, Kun and Park, Ju Yong and Kim, Se Hyun and Reddy, Pothala Reddi Sekhar and Materano, Monica and Mulaosmanovic, Halid and Mikolajick, Thomas and Jones, Jacob L and others},
  journal={Applied Physics Reviews},
  volume={8},
  number={2},
  pages={021312},
  year={2021},
  publisher={AIP Publishing LLC}
}

@article{pertsev2003coercive,
  title={Coercive field of ultrathin Pb (Zr 0.52 Ti 0.48) O 3 epitaxial films},
  author={Pertsev, NA and Rodr{\i}guez Contreras, J and Kukhar, VG and Hermanns, B and Kohlstedt, H and Waser, R},
  journal={Applied physics letters},
  volume={83},
  number={16},
  pages={3356--3358},
  year={2003},
  publisher={American Institute of Physics}
}

@article{kim2005critical,
  title={Critical thickness of ultrathin ferroelectric ba ti o 3 films},
  author={Kim, YS and Kim, DH and Kim, JD and Chang, YJ and Noh, TW and Kong, JH and Char, K and Park, YD and Bu, SD and Yoon, J-G and others},
  journal={Applied Physics Letters},
  volume={86},
  number={10},
  pages={102907},
  year={2005},
  publisher={American Institute of Physics}
}

@article{rollo2020stabilization,
  title={Stabilization of negative capacitance in ferroelectric capacitors with and without a metal interlayer},
  author={Rollo, T and Blanchini, F and Giordano, G and Specogna, R and Esseni, D},
  journal={Nanoscale},
  volume={12},
  number={10},
  pages={6121--6129},
  year={2020},
  publisher={Royal Society of Chemistry}
}

@article{cheema2022ultrathin,
  title={Ultrathin ferroic HfO2--ZrO2 superlattice gate stack for advanced transistors},
  author={Cheema, Suraj S and Shanker, Nirmaan and Wang, Li-Chen and Hsu, Cheng-Hsiang and Hsu, Shang-Lin and Liao, Yu-Hung and San Jose, Matthew and Gomez, Jorge and Chakraborty, Wriddhi and Li, Wenshen and others},
  journal={Nature},
  volume={604},
  number={7904},
  pages={65--71},
  year={2022},
  publisher={Nature Publishing Group UK London}
}

@article{li2018negative,
  title={Negative capacitance Ge PFETs for performance improvement: Impact of thickness of HfZrO x},
  author={Li, Jing and Zhou, Jiuren and Han, Genquan and Liu, Yan and Peng, Yue and Zhang, Jincheng and Sun, Qing-Qing and Zhang, David Wei and Hao, Yue},
  journal={IEEE Transactions on Electron Devices},
  volume={65},
  number={3},
  pages={1217--1222},
  year={2018},
  publisher={IEEE}
}

@inproceedings{zhou2016ferroelectric,
  title={Ferroelectric HfZrO x Ge and GeSn PMOSFETs with Sub-60 mV/decade subthreshold swing, negligible hysteresis, and improved I ds},
  author={Zhou, Jiuren and Han, Genquan and Li, Qinglong and Peng, Yue and Lu, Xiaoli and Zhang, Chunfu and Zhang, Jincheng and Sun, Qing-Qing and Zhang, David Wei and Hao, Yue},
  booktitle={2016 IEEE International Electron Devices Meeting (IEDM)},
  pages={12--2},
  year={2016},
  organization={IEEE}
}
\end{document}